\documentclass[twocolumn]{aastex63}
\usepackage[margin = 1in, includehead, footskip=0.25in]{geometry}
\usepackage{graphicx} 
\usepackage{natbib} 
\usepackage{amsmath}
\usepackage{xcolor}
\usepackage{soul}
\usepackage{bm}
\newcommand{\cuny}{Department of Physics, City University of New York Graduate Center, 365~5th~Ave, New~York, NY~10016, USA}
\newcommand{\queens}{Physics Department of Queens College, Queens College Science Building, 6530~Kissena~Blvd~B334, Queens, NY~11367, USA}

\newcommand{\pdot}{$\dot{P}$}
\newcommand{\ominusc}{$O-C$}
\newcommand{\sps}{\,s\,s$^{-1}$}

\begin{document}

\title{Gravitational Influence from Planets on the Measured Rates of Period Change\\ of Pulsating White Dwarfs}

\author{Ling Xuan Yao}
\affiliation{Lockheed Martin Corporation, 199 Borton Landing Rd, Moorsetown, NJ 08057, USA}
\author[0000-0002-0656-032X]{Keaton J.\ Bell}
\affiliation{\queens}
\affiliation{\cuny}

\author{Andrew Dublin}
\affiliation{\cuny}

\correspondingauthor{}\email{Keaton.Bell@qc.cuny.edu}

\begin{abstract}
The measured rates of period change, $\dot{P}$, in the signals of pulsating white dwarf stars are often interpreted as direct detections of structural changes from secular cooling. Due to the intrinsic nature of this quantity, $\dot{P}$ analysis has been used to probe fundamental physics, such as constraining the mass of hypothetical axion particles. However, most white dwarfs are expected to host planets that could induce an external source of period change, caused by the light travel time variations from reflex motion about the system barycenter. Assuming a plausible distribution of planets that could orbit white dwarf stars, we quantify the amount of reflex motion expected from undiscovered planets as an important source of extrinsic error in $\dot{P}$ analyses. While the median error from reflex motion is $\sim10^{-15}$\,s\,s$^{-1}$ (similar to the secular \pdot\ rates expected for cool DAV pulsators), individual close-in planets could cause $\dot{P}$ errors as large as $10^{-11}$\,s\,s$^{-1}$.
\end{abstract}


\section{Introduction}

White dwarfs are the compact final remnants of low- to intermediate-mass stars. The extreme temperatures and densities of white dwarf interiors make possible observational experiments that probe aspects of fundamental physics that are beyond the reach of terrestrial laboratories. Without substantial nuclear fusion, any energy radiated away in the form of photon and neutrino luminosities comes at the cost of the white dwarfs cooling and gravitationally contracting, modulo any additional sources of energy production or absorption (e.g., latent heat from core crystallization; \citealt{1968ApJ...151..227V}). The physics that affects the cooling rate of white dwarfs leaves signatures in the observed distribution of white dwarf luminosities \citep{2016NewAR..72....1G, 2022FrASS...9....6I,2019Natur.565..202T}.

\begin{table*}[t]
    \centering
    \begin{tabular}{|c c c c c|}
    \hline 
    Star & Type & Mode (s) &\pdot\ ($\text{s s}^{-1}$) & Reference \\
    \hline
      PG1159-035 & DOV & 516.1 & $(1.28\pm 0.03)\times 10^{-10}$  & \citet{Costa_2008}\\
      PG1351+489 & DBV & 489.3 & $(2.00\pm 0.90) \times 10^{-13}$   & \citet{rad}\\
      G117-B15A & DAV & 215.19 &$(5.12\pm 0.82)\times 10^{-15}$ & \citet{stableoptical} \\
      Ross 548 & DAV & 213.13 &$(3.30\pm1.10)\times 10^{-15}$  & \citet{R548} \\   
      L19-2 & DAV & 192.6 & $(3.00\pm0.60)\times 10^{-15}$ & \citet{C_rsico_2016} \\
    \hline
    \end{tabular}
    \caption{Measurements of \pdot\ for white dwarf pulsation modes from the literature. The type of pulsator is indicated. For white dwarfs with \pdot\ measured for multiple modes, only one was chosen here for reference.}
    \label{intro table}
\end{table*}

During certain phases of their cooling, white dwarf atmospheres meet the conditions to drive global, non-radial gravity-mode pulsations. White dwarfs may pulsate as they cool through three main instability strips, depending on their atmospheric compositions. From hottest to coolest these are the DOV (also known as GW Vir stars), DBV (helium-atmosphere), and DAV (hydrogen-atmosphere) pulsators. The pulsations resonate within the white dwarf interior and the oscillation frequencies are precisely tuned by the interior composition and structure. These pulsations can cause brightness variations at the white dwarf photospheres, enabling the direct measurement of white dwarf resonant frequencies via Fourier analysis of time-series photometry. The field of white dwarf asteroseismology aims to interpret these pulsation data to characterize the internal physics and structures of white dwarf stars \citep{2008ARA&A..46..157W,2019A&ARv..27....7C,2025arXiv250213258B}.

White dwarf interiors change subtly as they cool, and the resonant pulsation frequencies change with them, providing a direct opportunity to test relevant physical processes operating within individual stars. While some white dwarf pulsations are intrinsically incoherent---they vary in phase, frequency, and amplitude \citep[especially those with periods longer than $\approx$800\,s;][]{Hermes2017,Montgomery2020}---other white dwarf pulsations provide the most coherent optical signals in the known universe \citep{stableoptical}. A constant rate of period change of a stable pulsation signal, \pdot, can cause an observable drift in signal phase (compared to expectations for a perfectly coherent signal with a constant period). Because the phase drift compounds over time, a constant \pdot\ causes pulsation arrival times to drift as time squared, producing a parabolic trend in observed-minus-calculated (\ominusc) diagrams. Parabolic trends in \ominusc\ diagrams of white dwarf pulsations reveal \pdot\ values that are usually interpreted to be caused by the effects of secular cooling on the stellar structure \citep{1991ApJ...378L..45K}. 

Since hotter white dwarfs are cooling faster, faster secular \pdot\ values are expected for hotter classes of pulsating white dwarf. Measured \pdot\ values reported for the dominant pulsation modes of white dwarf stars are summarized in Table~\ref{intro table} (some are corrected for proper motion effects; see references for details). Measured \pdot\ rates are of-order $10^{-15}$\sps\ for cool DAVs \citep{stableoptical,R548,C_rsico_2016}, $\sim10^{-13}$\sps\ for the warmer DBV star PG1351+489 \citep{rad}, and as large as $\sim10^{-10}$\sps\ for the 516-second mode from hot DOV pulsator PG1159-035 \citep{Costa_2008}. Measurement of such subtle phase drift requires precise timing systems and accounting for Earth's motion about the Solar System barycenter \citep{1993BaltA...2..515K}.

The ability to infer white dwarf cooling rates from \pdot\ measurements provides a unique opportunity to constrain aspects of fundamental physics that are difficult to study elsewhere. \citet{2022FrASS...9....6I} summarizes four exotic aspects of fundamental physics that may affect rates of white dwarf pulsation period change: changes to the gravitational constant over time, $\dot{G}$ \citep{gdot, 2013JCAP...06..032C}; the magnetic dipole moments of neutrinos \citep{neutrino}; the masses of hypothetical axion particles \citep{1992ApJ...392L..23I,2008ApJ...675.1512B,2012MNRAS.424.2792C,2012JCAP...12..010C}; and the capture of weakly interacting massive particles \citep[WIMPs;][]{2018PhRvD..98j3023N}.

\ominusc\ analysis of white dwarf pulsations is also useful as a technique for detecting planets that orbiting planets. Planets cause reflex motion of the white dwarf about the center of mass, causing a periodic phase variation of the pulsation signal as the distance light has to travel from the star to an observer varies \citep{2018haex.bookE...6H}. Some white dwarf pulsations have been monitored for decades, providing upper limits on the presence of planets orbiting these particular targets along our line of sight \citep{mullally,2015ASPC..493..285W}. 

Approximately 97\% of stars in our galaxy will become white dwarfs at the end of their evolution \citep{2009ApJ...693..355W}. Exoplanet surveys suggest that most white-dwarf-progenitor stars host planetary systems \citep[e.g.,][]{exoplanet,2019AJ....158..109H}. As these stars evolve beyond the main sequence, close-in planets may be engulfed during the giant phase \citep{2009ApJ...705L..81V}, while most distant planets are expected to survive to the white dwarf stage \citep{Andryushin_2021}. The orbits of these planets expand adiabatically with mass loss of their host star, and then planet-planet interactions may cause the planets to migrate considerably to new orbital configurations \citep{2015MNRAS.447.1049V}. While intriguing planetary systems, planetary debris, and planet candidates have been detected around white dwarfs \citep{2003Sci...301..193S,2011ApJ...730L...9L,2015Natur.526..546V, 2019Sci...364...66M,2019Natur.576...61G,2020Natur.585..363V,2021ApJ...912..125G, 2021ApJ...917...41V,2021Natur.598..272B,2024ApJ...962L..32M, 2025ApJ...980...56H}, the demographics of distant planets orbiting white dwarfs are observationally underconstrained \citep{2024MNRAS.527.3532K}.

The expectation that planets are common around white dwarf stars should be taken into consideration when interpreting pulsation \pdot\ measurements. The signature of reflex motion from a companion in a circular orbit is a sinusoid in the \ominusc\ diagram. A long-period sinusoid could be misinterpreted as a parabola if only a fraction of the orbital period is observed. The extrinsic reflex motion effect that most pulsating white dwarfs likely exhibit could be mistaken for an intrinsic \pdot\ due to secular cooling of the white dwarf. Failing to account for the potential contribution of reflex motion to the \ominusc\ data on white dwarf pulsations could result in inaccurate conclusions about fundamental physics.

In this work, we consider the extent to which the gravitational influence from orbiting planets could systematically affects the measurement and interpretation of \pdot\ from pulsating white dwarf stars. We aim to quantify the amount of extrinsic error that should be considered in addition to the intrinsic statistical error from fitting a parabola to \ominusc\ data. 
Section~\ref{sec:two} characterizes the parabolic term of the Taylor series expansion of sinusoidal \ominusc\ signals caused by planets in circular orbits. We adopt a plausible model \citep[based on][]{Andryushin_2021} for the distribution of planets suspected to orbit most white dwarf stars to determine the expected distribution of apparent \pdot\ caused by reflex motion in Section~\ref{sec:system}. We find that the expected errors from reflex motion span many orders of magnitude, with a median value of $\sim10^{-15}$\,s\,s$^{-1}$ (the same order of magnitude as the expected evolutionary \pdot\ values observed in the coolest DAV pulsators), but with some as large as $\sim10^{-11}$\,s\,s$^{-1}$.

\section{Planetary effects on \pdot\ measurements}\label{sec:two}

An important tool for analyzing the coherence of pulsation signals is the Observed$-$Calculated (\ominusc) diagram. This diagram records the measured difference in arrival time between an observed signal and the timing expected for a strictly periodic signal. If a pulsation signal with initial reference period $P_0$ is assumed to change at a constant rate, \pdot, we expect to see a parabolic trend in the \ominusc\ diagram \citep[Section 9.6]{textbook} following the relation
\begin{equation} \label{eq:parabolic}
    O-C = \frac{1}{2}P_0\dot{P}E^2,
\end{equation}
where $E$ represents the number of elapsed cycles of the reference pulsation period, or epochs, as a proxy for time.

On the other hand, a signal under the influence of circular reflex motion exhibits a sinusoidal \ominusc\ trend as the white dwarf orbits the system barycenter. In the limit of only observing a small fraction of an orbital period, the \ominusc\ diagram can look identical to a parabola within the measurement noise, leading to a potential misinterpretation as an evolutionary \pdot. We quantify the parabolic content of sinusoidal reflex motion in this limit with a Taylor expansion. We interpret the \pdot\ that would be inferred from reflex motion to represent a source of extrinsic error that undiscovered planets introduce to attempts to measure intrinsic \pdot\ values.

\subsection{Analytical derivation}\label{sec:analytic}

\begin{figure}[t]
    \centering
     \includegraphics[width=5.5cm]{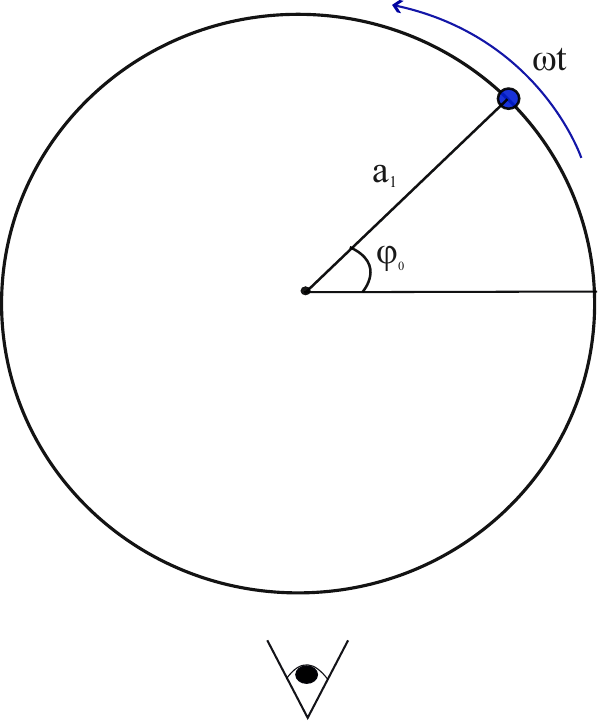}
    \caption{Diagram of a pulsating white dwarf (blue dot) located at phase angle $\phi_0$ as it orbits the center of mass (black dot) at distance $a_1$ with angular velocity $\omega$. The eye at bottom indicates the direction of the observer. The time it takes for the signal to be received varies with the distance to the observer.
    }
    \label{fig:schematic}
\end{figure}

For a white dwarf of mass $m_1$ orbited by a planet of mass $m_2$ at a constant separation $a$ in a circular orbit, the radius of the reflex motion of the white dwarf is its distance from the center of mass, $a_1=a m_2/(m_1+m_2)$.
Figure~\ref{fig:schematic} depicts the scenario where a pulsating white dwarf orbits the center of mass with angular velocity $\omega$ at distance $a_1$; at time $t=0$, the white dwarf is positioned in its orbit at phase angle $\phi_0$. Compared to a signal originating from the center of mass (if there were no planet), there is a lag of $(a_1/c)\sin{\phi_0}$ in the time that the pulsation signal arrives to the observer because of the extra distance the signal must travel. When viewing the orbit from inclination angle $i$, the observed pulsation times vary sinusoidally with semi-amplitude $(a_1/c)\sin{i}$ and angular frequency $\omega$. 

To track how reflex motion modulates the phase of the observed pulsation signal, we could record, for example, the times of observed brightness maxima. If we had a constant pulsation period, $P_0$, then relative to one reference time of pulsation maximum, $t_0$, future epochs of pulsation maxima are expected to occur integer pulsation periods later, such that we can calculate future times of pulse maximum $C(E) = t_0+P_0E$, for integer epoch number $E$ \citep{textbook}.
These calculated times are the $C$ in \ominusc. Because of reflex motion, the observed times of pulse maximum depart from this expectation sinusoidally as
\begin{equation} \label{eq:oc}
\begin{split}
    O-C  = \frac{a_1}{c}\sin(\omega t + \phi_0)\bm{\sin i} \\
    = \frac{a_1}{c}\sin(\omega P_0E+\phi_0)\bm{\sin i},
\end{split}
\end{equation}
where the white dwarf starts at phase angle $\phi_0$ at time $t=0$ such that $\phi=\omega t+\phi_0$, and we substitute for time the number of elapsed pulsation cycles $t = P_0E$.

Using the trigonometry identity for the sine of a sum of angles and Taylor expanding, we get
\begin{equation}\label{eq:taylor}
\begin{split}
    O-C=\frac{a_1}{c}\sin{i}\biggl[\left(\omega P_0E-\frac{\omega^3P_0^3E^3}{3!}+\ldots\right)\cos(\phi_0)\\
    +\left(1 - \frac{\omega^2P_0^2E^2}{2!} + \ldots\right)\sin(\phi_0)\biggr].
\end{split}
\end{equation}
The even powers of the expansion depend on the orbital position of the white dwarf along our line of sight, and the odd powers depend on the component perpendicular to our line of sight.

The linear term in Eq.~\ref{eq:taylor} represents the difference between the intrinsic pulsation period and the expectation value for the best-fit pulsation period to the full data set \citep{textbook}. The implications for how reflex motion can systematically Doppler shift white dwarf pulsation period measurements themselves are explored in Appendix~\ref{app:periodoffsets}. Since calculated times for expected pulse arrivals are based on the best-fit period, the linear term is typically calibrated out of \ominusc\ diagrams.
Terms of higher power are increasingly difficult to detect in the data, as the orbital period is much longer than the pulsation period ($\omega P_0 \ll 1$). 
In the limit of observing a small fraction of an orbit ($\omega t \ll 1$), \ominusc\ records of reflex motion can be statistically consistent with a parabola within measurement errors. 

If the parabolic \ominusc\ trend caused by reflex motion is misinterpreted to be caused by an evolutionary rate of period change, we could equate the quadratic ($E^2$) term in Eq.~\ref{eq:taylor} to Eq.~\ref{eq:parabolic} to infer
\begin{equation} \label{eq:squaredterm}
    \dot{P} = -\frac{a_1P_0\omega^2}{c}\sin{\phi_0}\sin{i}.
\end{equation} 
The parabolic trend is strongest when the white dwarf is observed at its nearest or furthest point along our line of sight. 
Substituting from Kepler's third law, we get
\begin{equation} \label{eq:physical}
    \dot{P}  = -\frac{Gm_2}{a^2}\frac{P_0}{c}\sin{\phi_0}\sin{i},
\end{equation}
independent of the mass of the star.
Written this way, it is insightful to note that the apparent \pdot\ signal is proportional to the pulsation period times the component of the star's acceleration vector along our line of sight, supporting the interpretation that the reflex motion effects in an \ominusc\ diagram are caused by the varying Doppler shift of the pulsation signal. This result agrees with the expression presented in \citet[eq.~10]{1991ApJ...378L..45K}.

Equation~\ref{eq:physical} can be rearranged to further examine the contributions of individual terms on the upper limit for inferred \pdot\ (for $|\sin\phi_0\sin i|=1$):

\begin{equation}\label{eq:loggedO-C}
\begin{split}
    \log_{10}\left|\dot{P}\right| \leq \log_{10}\left(\frac{Gm_2P_0}{a^2c}\right)\\
    = \log_{10}\left(\frac{(215\,{\rm s})G{\rm m_\oplus}}{c\,{\rm a_\oplus}^2}\right)+\log_{10}\left(\frac{m_2}{\rm m_\oplus}\right)\\
    -2\log_{10}\left(\frac{a}{\rm a_\oplus}\right)+\log_{10}\left(\frac{P_0}{215 \,{\rm s}}\right)\\
    = -13.89 + \log_{10}\left(\frac{m_2}{\rm m_\oplus}\right)-2\log_{10}
    \left(\frac{a}{\rm a_\oplus}\right)\\
    +\log_{10}\left(\frac{P_0}{215\,{\rm s}}\right),
\end{split}
\end{equation}
where a$_\oplus = 1$\,AU, and the pulsation period is scaled to 215 seconds, roughly the pulsation period of G117-B15A. Compared to this expression for the maximum effect, actual inferred \pdot\ values will be reduced by observed orbital phases and inclinations.
 
Using Kepler's third law, this upper limit can be reparameterized in terms of the planet and star masses ($m_2$ and $m_1$) and the orbital period ($P_{\rm orb}$) as
\begin{equation}\label{eq:loggedphys}
\begin{split}
    \log_{10}\left|\dot{P}\right| \leq -13.89 +\log_{10}\left(\frac{m_2}{\rm \rm{m}_\oplus}\right) \\
    -\frac{2}{3}\log_{10}\left(\frac{m_1+m_2}{\rm{M}_\odot}\right)-\frac{4}{3}\log_{10}\left(\frac{P_{\text{orb}}}{1\,\text{yr}}\right)\\
    +\log_{10}\left(\frac{P_0}{215\,{\rm s}}\right).
\end{split}
\end{equation}

\subsection{Validation with simulated data}\label{sec:simulated}

To test the validity of our analytic result, we infer \pdot\ values from \ominusc\ diagrams for simulated  light curves of pulsating white dwarfs exhibiting circular reflex motion. We explore here a case that would be consistent with the value $\dot{P} = 5.47 \times 10^{-15}$\sps\ determined from over 50 years of \ominusc\ data on G117-B15A, prior to applying a proper motion correction \citep{stableoptical}. For the pulsation period of 215 seconds, Eq.~\ref{eq:physical} implies that a Jupiter-mass planet orbiting edge-on with a separation of 27.25\,AU could mimic this \pdot\ if the average acceleration of the white dwarf points directly away from the observer \citep{stableoptical}. We simulate individual days of time series observations spanning 50 years using a canonical value of 0.6\,$\rm{M}_\odot$ for the pulsating white dwarf, and choosing pulsation amplitudes, noise, and observational revisit times that produce an \ominusc\ diagram with a plausible signal-to-noise ratio. Each day's light curve is analyzed with the {\sc Pyriod}\footnote{\url{https://github.com/keatonb/Pyriod}} frequency analysis code, which fits sinusoidal signals to the time series data. The \ominusc\ diagram is constructed by fitting a constant-period sinusoid to each day's light curve and recording variations in measured signal phase \citep{2013ASPC..469...45D, 2014MNRAS.441.2515M}. A preliminary \ominusc\ diagram is constructed with an approximate reference period, and then the linear term of a parabolic fit is used to refine the reference period \citep[Section 9.6]{textbook}. Our code to simulate and analyze these data sets is available online.\footnote{\url{https://github.com/YaoLing-boop/OC-sim}}

\begin{figure}[]
    \centering
    \includegraphics[width=7cm]{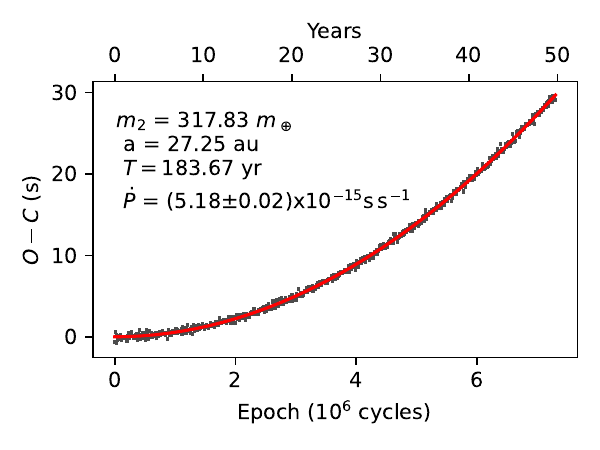}
    \caption{Simulated \ominusc\ diagram of a 0.6\,$\rm{M}_\odot$ pulsating white dwarf with a Jupiter-mass planet orbiting edge-on at 27.25\,AU, with average orbital position $\phi =\frac{3\pi}{2}$ (Figure~\ref{fig:schematic}). The data are simulated with 60-second observing cadence for 1 day every month for 50 years. 
    Eq.~\ref{eq:physical} predicts a value  $\dot{P} = (5.47 \pm 0.82) \times 10^{-15}$\,s\,s$^{-1}$ in the Taylor limit consistent with that measured for G117-B15A \citep{stableoptical}. See text for discussion.}
    \label{fig:O-C}
\end{figure}

Figure~\ref{fig:O-C} displays the \ominusc\ diagram simulated with physical reflex motion meant to mimic the G117-B15A signal. While the observed period change is caused by a sinusoidal reflex motion, the data are statistically consistent (reduced Chi-squared $\chi^2_{\rm red} = 1.0$) with the best-fit parabola, which could represent a constant rate of period change, \pdot. From fitting a parabola to these 50 years of simulated data, we would infer a value of $\dot{P} = (5.18\pm0.02)\times10^{-15}$\sps, which is close to, but slightly lower than, the targeted value of $\dot{P} = 5.47 \times 10^{-15}$\,\sps. The discrepancy results from 50 years being a significant fraction (27\%) of the 183.7-year orbital period. While this departs from the low-order Taylor limit, where the star is assumed to have a constant acceleration vector, the \ominusc\ data still appear parabolic, and the Taylor expansion gives a reasonable estimate of the magnitude of inferred \pdot. As we fit to shorter subsets of the \ominusc\ data, we confirm that the inferred \pdot\ approaches the value predicted by Eq.~\ref{eq:physical}. We describe how the inferred \pdot\ value departs from expectations from the Taylor expansion with longer observing duration in Appendix~\ref{app:duration}.

\subsection{Probability distribution for \pdot\ from individual planets}\label{sec:pdotdist}

Section~\ref{sec:analytic} presented an analytical result for the \pdot\ that would be inferred from observing a pulsating star exhibiting circular reflex motion due to a companion of given mass and separation, giving a maximum effect described in Eq.~\ref{eq:loggedO-C}. The observed effect is proportional to the component of the star's acceleration vector along our line of sight (Eq.~\ref{eq:physical}). For an undiscovered planet, the acceleration vector is randomly oriented, and the probability distribution for \pdot\ inferred from a hypothetical planet of given mass and separation is a uniform density function between $\pm\frac{Gm_2P_0}{a^2c}$.

\begin{figure}
    \centering
    \includegraphics[width=7.5cm] {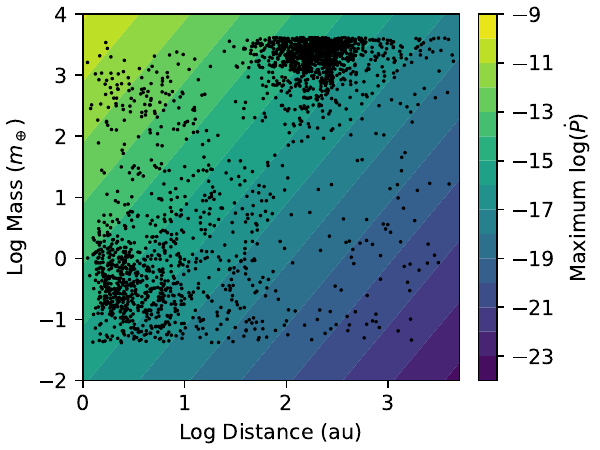}
    \caption{Contour plot showing the maximum $\log_{10}(\dot{P})$ values for various combinations of orbital separation and planetary mass for a 215-second pulsation period. The black dots show the final masses and separations of planets simulated to survive evolution of their host stars to the white dwarf phase by \citet{Andryushin_2021}.}
    \label{fig:contour}
\end{figure}

Figure \ref{fig:contour} displays contours for the quantity $\frac{Gm_2P_0}{a^2c}$ that sets the width of the isotropic \pdot\ distribution for a reference pulsation period $P_0 = 215$\,s. The reference pulsation period of 215\,s matches the signal from G117-B15A that has been monitored for decades \citep{stableoptical}, and the width of the \pdot\ distribution scales as $P_0/(215$\,s) for different pulsation periods. Figure~\ref{fig:contour} shows that larger \pdot\ values could be inferred for planets that are more massive or that orbit closer to the white dwarf, which could cause a greater acceleration of the pulsating star along our line of sight. The final distances and masses of planets that are simulated to survive the evolution of their host stars to the white dwarf phase by \citet{Andryushin_2021} are shown as black dots for reference. Actual inferred \pdot\ values will depend on the orientation of the star's acceleration vector during the observations.

\section{Average \pdot\ in a System of Planets}\label{sec:system}

By evaluating the net \pdot\ that would be inferred from many hypothetical but plausible systems of planets, we can estimate the magnitude of extrinsic error that undiscovered planets are expected to introduce to measurements of secular \pdot.
We adopt the population of planets that were simulated to survive stellar evolution to the white dwarf phase by \citet{Andryushin_2021}.
These models are suitable for estimating the magnitude of extrinsic \pdot\ error in the absence of strong observational constraints.
All of our calculations assume a reference pulsation period of 215\,s, and resulting \pdot\ distributions would scale as $P_0/(215$\,s) for different mode periods.

The planet population synthesis work from \citet{Andryushin_2021} starts with initial planet masses and separations based on the models of \citet{Alibert_2013}, and then simulates how these orbits evolve due to stellar mass loss as simulated with the MESA stellar evolution code \citep{MESA}. Stars with initial masses of 1--8\,$\rm{M}_\odot$ were evolved to the white dwarf stage, with the semi-major axes of planets calculated and updated at each step \citep{Andryushin_2021}. For simplicity, inter-planetary interactions and tidal forces were not considered, though these interactions could have a significant effect on planetary migration \citep{2015MNRAS.447.1049V}.

\begin{figure}[t]
    \centering
    \includegraphics[width=7cm]{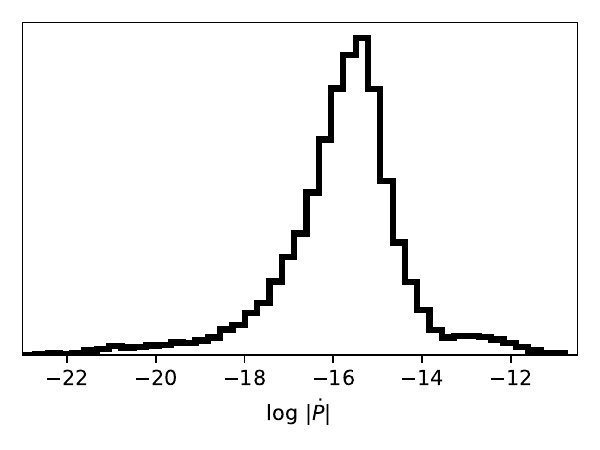}
    \caption{Histogram of $\log_{10}|\dot{P} |$ values expected from individual planets simulated to survive stellar evolution to the white dwarf stage by \citet{Andryushin_2021}. The \pdot\ for each planet was calculated with Eq.~\ref{eq:physical}  using a 215-second reference period for random inclination and orbital phase.} 
    \label{single}
\end{figure}

To assess the distribution of inferred \pdot\ values that could be caused by individual, plausible planets that could orbit white dwarf stars, we select the masses and distances of random planets simulated to survive the evolution of their host star to the white dwarf stage by \citet{Andryushin_2021}. Each planet is assigned a random orbital phase ($\phi$ drawn from a uniform distribution between 0 and 2$\pi$) and inclination (for isotropic orientation with $\cos{i}$ drawn from a uniform distribution between 0 and 1). \pdot\ values are calculated from Eq.~\ref{eq:physical}. Figure~\ref{single} shows a histogram of the base-10 logarithm of the absolute value of \pdot\ caused by random, plausible planets. Induced \pdot\ values span many orders of magnitude. The most massive, close-in planets, with acceleration vectors closely aligned with our line of sight can cause \pdot\ values exceeding $10^{-12}$\,\sps. The distribution peaks strongly near the median value of $\log{|\dot{P}|} \approx -15.7$. The sign of \pdot\ will be positive or negative depending on whether the star is observed on the near or far side of its orbit.

\begin{figure}[t]
    \centering
    \includegraphics[width=7cm]{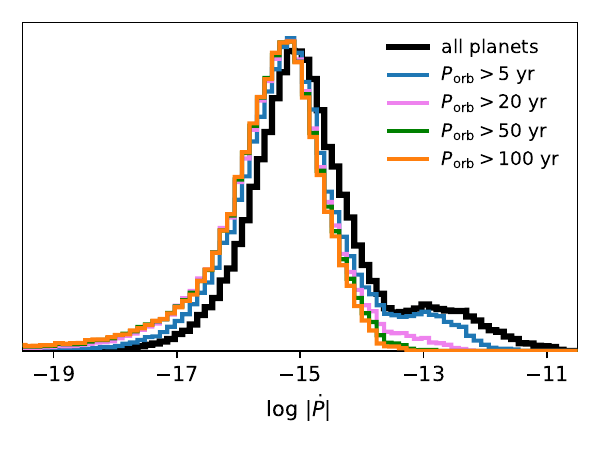}
    \caption{Histogram of $\log_{10}{|\dot{P}|}$ values expected from systems of three coplanar planets. The black distribution represents the effects of all combinations of planets. The other distributions exclude contributions from planets with orbital periods shorter than the indicated cutoff times. These planets may be ruled out from contributing to parabolic \ominusc\ trends if the observing duration is long enough that reflex motion would no longer appear consistent with a parabola.}
    \label{fig:3planets}
\end{figure}

In reality, most white dwarfs could be host to multiple planets, each of which will contribute separately to the net acceleration of a star during observations. In this case, we can approximate the inferred \pdot\ as the sum of contributions from each planet. In some cases planets with similar phases will combine constructively to induce a greater apparent \pdot, while planets on opposite sides of their orbits will decrease the inferred \pdot. Following the assumptions and results of \citet{Andryushin_2021}, if planetary systems begin with eight planets on average and around 60\% are engulfed with negligible ejections, we might expect white dwarfs to host three planets on average. While we expect that properties of planets orbiting white dwarfs are likely correlated through the planet formation and migration process \citep[e.g.,][]{2019MNRAS.485..541C}, for the purposes of evaluating the order-of-magnitude of extrinsic \pdot\ error, we consider simplified multi-planet systems consisting of three planets with masses and semi-major axes that survived the \citeauthor{Andryushin_2021} simulations. These planets are given random initial positions in circular orbits, but are assumed to be coplanar with a shared random inclination. 

The histogram of the base-10 logarithm of the absolute inferred \pdot\ expected from different simulated three-planet systems is displayed in black in Figure~\ref{fig:3planets}. The distribution is bimodal in log space, with a median value for $\log{|\dot{P}|}$ of $-15.0$, but with 12.4\% of systems in the high-\pdot\ secondary peak of the distribution with $\log{|\dot{P}|} > -13.5$ caused by the contributions of individual massive planets on relatively close orbits.

\begin{figure}[t]
    \centering
    \includegraphics[width=7.5cm]{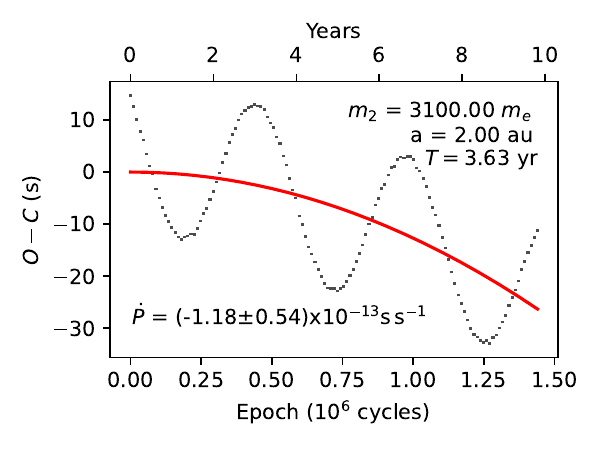}
    \caption{Simulated \ominusc\ diagram for circular reflex motion of a pulsating white dwarf monitored for multiple orbital periods. The system contains a $0.6\,{\rm M}_\odot$ white dwarf with a 215-second pulsation period orbited edge-on by a $3100\,{\rm M}_\oplus$ planet at a distance of 2.0 AU. The white dwarf has average orbital position $\phi = \frac{3\pi}{2}$. The \ominusc\ data are inconsistent with a parabola, and the \pdot\ value that would be inferred from a parabolic fit has the has the opposite sign and is two orders of magnitude smaller than expected from Eq.~\ref{eq:physical} in the Taylor limit.}
    \label{fig:bigfig}
\end{figure}

Considering that tracking phase changes of pulsation signals is an exoplanet discovery method \citep{2018haex.bookE...6H}, large extrinsic error to \ominusc\ analyses from planets with short-period orbits becomes less of a concern when the systems are monitored for long observing durations. If a system is observed through most of the orbital period of a planet, the \ominusc\ trend will either depart from a parabola, revealing its orbital nature, or the contribution to the parabolic term will become less significant (see Appendix~\ref{app:duration}). To demonstrate this, Figure~\ref{fig:bigfig} shows an \ominusc\ diagram simulated following the procedure of Section~\ref{sec:simulated}. The plot shows ten years of data for a 3100\,m$_\oplus$ planet (near the high-mass limit of the \citealt{Andryushin_2021} population) simulated to orbit a canonical 0.6\,M$_\odot$ white dwarf at a distance of 2 AU, corresponding to an orbital period of 3.63\,yr. The orbit is observed with edge-on orientation, and the average orbital phase at the midpoint of observations is $\phi=3\pi/2$. The massive planet causes large sinusoidal variations that clearly depart from the best-fit parabola.  Eq.~\ref{eq:physical} predicts the inferred \pdot\ from a parabolic fit in the short-observing-duration limit of $\sim10^{-11}$\,\sps, while the \pdot\ value that would be inferred from the fit in Fig.~\ref{fig:bigfig} the opposite sign and is two orders of magnitude smaller ($\sim-10^{-13}$\,\sps ).

The longer a system is monitored, the more we can rule out the influence of hypothetical short-period planets that could induce the strongest accelerations. Figure~\ref{fig:3planets} shows how the probability distribution for expected extrinsic error on \pdot\ changes as we reject the contributions from planets with increasingly long orbital periods, for an assumed white dwarf mass of 0.6\,$\rm{M}_\odot$. If we can rule out the influence of planets with orbits shorter than 50 years, for example, the higher-\pdot\ component of the original bimodal distribution disappears. In this case the largest expected \pdot\ errors have $\log{|\dot{P}|} \approx -13$ and the median \pdot\ error converges toward $\log{|\dot{P}|} \approx -15.5$.
The main 215-second pulsation mode of G117-B15A has been monitored since 1974 \citep{stableoptical}, allowing us to relax our concerns that the parabolic \ominusc\ trend is caused by planets with orbits shorter than $\approx100$ years.

\section{Discussion and Conclusions}

Most white dwarf stars are expected to host exo\-planetary systems. While these exoplanets may be difficult to detect, their gravitational influence on pulsating white dwarfs can cause variations in measured pulsation signal phases as the distance light must travel to our telescopes is modulated by reflex motion \citep{mullally}. When only a small fraction of an orbit is observed, these variations can appear parabolic in an \ominusc\ diagram, consistent with expectations for a pulsation mode exhibiting an constant intrinsic rate of period change. Measured rates of white dwarf pulsation period change can be used to probe fundamental physics \citep{2022FrASS...9....6I}, and the high potential impact of such studies warrants careful consideration of sources of systematic error. In this work, we have shown that the magnitude of extrinsic error that plausible systems of planets orbiting white dwarf stars introduce to efforts to measure evolutionary \pdot\ can often exceed the intrinsic values. This represents a challenge for accurately interpreting pulsating white dwarf \ominusc\ diagrams.

The magnitude of a \pdot\ signal that can be mimicked by reflex motion is proportional to the product of the pulsation period and the component of the star's acceleration vector along our line of sight. Constructing plausible three-planet exoplanetary systems based on the results of \citet{Andryushin_2021}, we find a distribution of induced \pdot\ values that spans many orders of magnitude. The median error for a 215-s pulsation period is expected to be $\log{|\dot{P}|} \approx -15.0$, consistent with the rates of period change observed from hydrogen-atmosphere pulsating white dwarf stars belonging to the DAV class. However, individual planetary systems can induce signals as large as $\log{|\dot{P}|} \approx -11$, while effects of reflex motion in other systems could be orders of magnitude smaller than evolutionary \pdot.

The \citeauthor{Andryushin_2021} models of planet migration during stellar evolution do not account for planet-planet scattering, which could redistribute planets within a system \citep{2015MNRAS.447.1049V}, including to smaller orbits where they would cause larger \pdot\ errors. The actual extrinsic error on \pdot\ from reflex motion depends on the true distribution of planets orbiting white dwarfs. In the absence of strong observational constraints, \citeauthor{Andryushin_2021} provide a reasonable model of planet demographics to assume for estimating the order of magnitude of expected extrinsic errors. Our calculations assume circular orbits for simplicity; elliptical orbits will tend to broaden the inferred \pdot\ distribution as gravitational acceleration is stronger/weaker near periapsis/apoapsis. For long observing campaigns, we can rule out the influence of close-in planets with short orbital periods, which could contribute the largest errors to the interpretation of \ominusc\ data.

The hotter classes of pulsating white dwarf are cooling much faster, with expected intrinsic \pdot\ values of order $10^{-13}$\,s\,s$^{-1}$ for helium-atmosphere pulsating DBVs, and $\sim10^{-10}$\,s\,s$^{-1}$ for hot DOVs. Planetary reflex motion will cause a smaller extrinsic \textit{fractional} error for these classes of pulsator, especially for planetary systems that have not had as much time to potentially scatter planets to smaller orbits \citep{2015MNRAS.447.1049V}.

Reflex motion Doppler shifts pulsation signals, and we show in Appendix~\ref{app:periodoffsets} that the expected extrinsic error on pulsation period measurements themselves can be as large as $\sim 10^{-4}$\,s for a 215-s pulsation mode of a 0.6\,M$_\odot$\ white dwarf. This is smaller than the typical intrinsic uncertainties on pulsation period measurements of $\sim10^{-3}$\,s \citep[e.g.,][]{Hermes2017}, and therefore not likely to have significant consequences for white dwarf asteroseismology. A greater contribution to period uncertainty from Doppler shifts will come from their overall radial velocity through the galaxy \citep[radial velocity dispersion $\sim30$\,km\,s$^{-1}$][]{2017MNRAS.469.2102A}, or for those in stellar binary systems. A larger contribution to systematic error for white dwarf asteroseismology comes from the physics used to model white dwarfs \citep{2022FrASS...9.9045G}.

The extrinsic contribution to \pdot\ from reflex motion can be positive or negative. For an individual pulsating white dwarf, all pulsation modes will experience the same time delays from reflex motion \citep{2013ApJ...765....5D}, while intrinsic rates of period change can vary between modes \citep[e.g.,][]{2004A&A...428..159C}. When interpreting measurements of pulsating white dwarfs, one should consider that there is a random effect on the system overall, producing a systematic correlated error on all observed modes. If long-term monitoring can be achieved for a large population of pulsating white dwarfs, errors on individual systems from gravitational effects will tend to average out for the ensemble, which could lead to more robust conclusions about fundamental physics and cosmology. Already, it is intriguing that the \pdot\ measurements for the main modes of all five published white dwarfs trend positively (Table~\ref{intro table}), though this could be influenced by publication bias. Considering all of the modes of the DOV pulsator PG\,1159-035, they show a mix of positive and negative \pdot\ values \citep{Costa_2008}, while seismic models predict \pdot\ for all modes should be positive \citep{2022ApJ...936..187O}. \pdot\ for some PG\,1159-035 modes appear to have changed sign since the start of monitoring in 1983 \citep{1985ApJ...292..606W}, suggesting an additional physical effect with a timescale of years \citep{2022ApJ...936..187O}.

There may be white dwarfs that do not host planets. In other systems, all planets will be inclined near $0^\circ$ so that there is no component of reflex motion along our line of sight. In these systems, \pdot\ measurements could directly trace evolutionary processes. Other observational evidence could inform us about the likelihood of planets contributing to \ominusc\ trends in individual systems. In the best-studied case of G117-B15A, \citet{stableoptical} argue that the presence of distant planets is limited by the proximity of the binary companion G117-B15B. If G117-B15A and B are currently observed at maximum separation, planets beyond $\approx240$\,AU are not expected to maintain stable orbits \citep{2005A&A...434..355M}; however, greater planet separations are possible if G117-B15A and B are not currently near maximum separation. Meanwhile, the \ominusc\ data for G117-B15A already rule out the presence of close-in planets orbiting along our line of sight in this system out to a Saturn at Saturn's distance from the Sun \citep{mullally}.

\acknowledgements We thank the referee for feedback that helped to improve the paper. We thank that authors of \citet{Andryushin_2021} for sharing detailed results from their study. This material is based upon work supported by the National Science Foundation under Grant
No.\ AST-2406917. An early version of this work appears in a Masters thesis \citep{msthesis}. Thanks to Joshua Tan for providing feedback as a member of the thesis committee. The views and content presented in this document are not endorsed by Lockheed Martin Corporation and were not prepared as part of LXY's employment with the company.

\appendix

\vspace{-5mm}
\section{Extrinsic Error on Pulsation Period Measurements} \label{app:periodoffsets}

Motion of a pulsating star along our line of sight will Doppler shift the observed pulsation periods from the intrinsic periods, which could affect their asteroseismic interpretation. In the non-relativistic limit, for a star moving with a radial component of velocity $v_r$, the Doppler shift of the measured period from the intrinsic period can be expressed as
\begin{equation}\label{eq:doppler}
   \Delta P_0= P_0^\prime - P_0 = \frac{v_r}{c}P_0,
\end{equation}
where $c$ is the speed of light.
The radial component of the velocity vector of the star is (see Figure~\ref{fig:schematic})
\begin{equation} \label{eq:radial}
    v_r = v\cos({\phi}_0)\sin{(i}) = a_1\, \omega \cos({\phi_0})\sin({i}).
\end{equation}
The $\sin({i})$ factor is included to account for orbital inclination and $a_1$ is the distance of the star from the center of mass. Substituting Eq.~\ref{eq:radial} into Eq.~\ref{eq:doppler},
\begin{equation}
    \Delta P_0 = \frac{a_1}{c}P_0\omega \cos(\phi_0)\sin(i).
\end{equation}
This matches the coefficient on the linear term in the Taylor expansion of a reflex-motion \ominusc\ diagram given in Eq.~\ref{eq:taylor}, demonstrating that this difference between intrinsic and expected best-fit pulsation period is caused by Doppler shift. The linear term in an \ominusc\ diagram represents a discrepancy between the best-fit period to the dataset and the reference period \citep[used to calculate expected pulsation times, $C$;][]{textbook}. In the Taylor expansion for circular reflex motion in an \ominusc\ diagram presented in Section~\ref{sec:analytic}, we take the intrinsic pulsation period to be the reference period. The coefficient on the linear term in Eq.~\ref{eq:taylor} therefore represents the difference between the best-fit reference period ($P_0^\prime$) and the true intrinsic pulsation period. Similar to our concern for how orbital motion could affect \pdot\ inferences, the period measurements themselves can have systematic errors because of reflex motion. 

Substituting from Kepler's third law for the masses of the pulsating star and its orbital companion, $m_1$ and $m_2$, and their separation, $a$, we get
\begin{equation} \label{eq:deltaP}
    \Delta P_0 = \frac{P_0m_2}{c}\sqrt{\frac{G}{a(m_1+m_2)}}\cos{\phi_0}\sin{i},
\end{equation}
Similar to Eq.~\ref{eq:loggedO-C}, we can express the upper limit on the Doppler period offset (for $|\cos\phi_0\sin i|=1$) as
\begin{equation}
    \log_{10}\left|\Delta P_0\right|\leq -7.19+ \log_{10}\left(\frac{m_2}{\rm \rm{m}_\oplus}\right) - \frac{1}{2}\log_{10}\left(\frac{m_1+m_2}{\rm M_\odot}\right) - \frac{1}{2}\log_{10}\left(\frac{a}{\rm a_\oplus}\right)+\log_{10}\left(\frac{P_0}{215\,{\rm s}}\right)
\end{equation}

\begin{figure}[t]
    \centering
    \includegraphics[width=8cm]{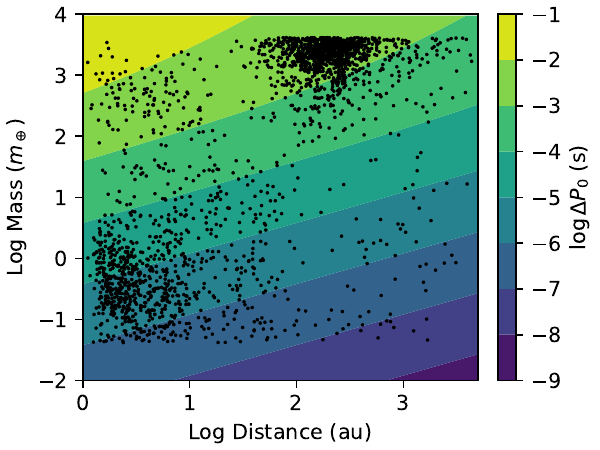}
    \caption{Contour plot showing the log of the maximum difference between an intrinsic 215-second pulsation period and the observed period, $\log_{10}(\Delta P_0)$, due to Doppler shift caused by reflex motion of a 0.6$M_\odot$ white dwarf moving along our line of sight from planets of various masses and distance. The black dots show the final masses and separations of planets simulated to survive evolution of their host stars to the white dwarf phase by \citet{Andryushin_2021}.}
    \label{varyingA}
\end{figure}

Figure~\ref{varyingA} displays contours for the maximum magnitude of difference expected between intrinsic and measured pulsation period for a 215-second mode caused by Doppler shift of a canonical 0.6\,M$_\odot$ white dwarf moving directly along our line of sight for different planetary companion masses and separations. These values assume that the system is observed for a small fraction of an orbital period. Actual systematic period errors scale with $P_0/(215\,{\rm s})$ and will depend on the white dwarf mass and its direction of motion relative to our line of sight.

\begin{figure}[t]
    \centering
    \includegraphics[width=0.4\linewidth]{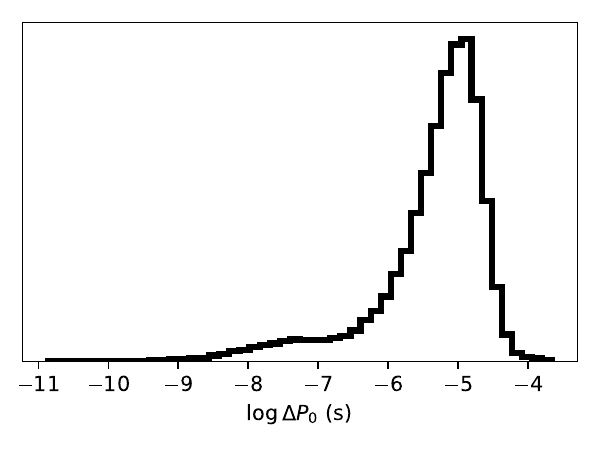}
    \caption{Distribution of the base-10 log of the difference between measured and intrinsic period due to Doppler shift (Eq.~\ref{eq:deltaP}) from reflex motion in 100{,}000 simulated three-planet systems drawn from the results of \citet{Andryushin_2021}. Planets in each system are assigned random orbital phases in coplanar orbits.}
    \label{fig:perdiff_dist}
\end{figure}

To assess the magnitude of extrinsic error we expect reflex motion to introduce to the measurement of a 215-second mode period, we simulate plausible planetary systems by drawing combinations of three planet masses and separations from the \citet{Andryushin_2021} models following the methodology of Section~\ref{sec:system}. For these systems, the net $\Delta P_0$ is calculated as the sum of $\Delta P_0$ values caused by each individual planet. The distribution of resulting $\Delta P_0$ values is plotted in Figure \ref{fig:perdiff_dist}. The mean of the distribution is centered at $\Delta P_0 = 10^{-5.4}$\,s, and it extends up to $\Delta P_0 \approx 10^{-4}$\,s.

\section{Inferred \pdot\ dependence on observing duration}\label{app:duration}

\begin{figure}[b]
    \centering
    \includegraphics[width=8cm]{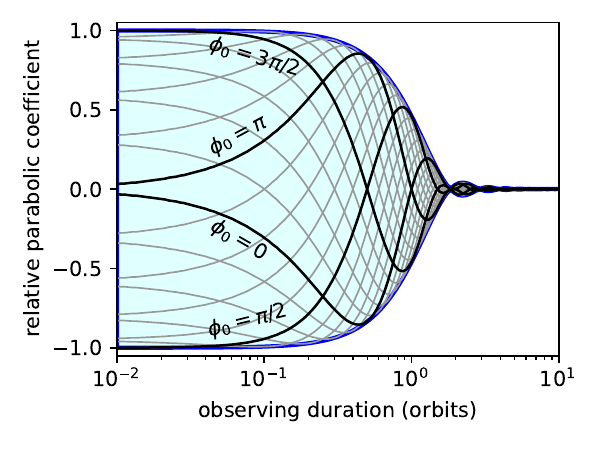}
    \caption{The relative best-fit coefficient of the squared term of a parabola to a sinusoid evolves with observing duration, in terms of the number of full cycles (orbits) observed. 
    Coefficients are scaled relative to the maximum squared-term coefficient that would fit to the sinusoid extrema in the limit of short observing duration.
    The gray curves show how this best-fit coefficient evolves as more data are collected for different starting phases of the sinusoid. The black curves represent starting phases $\phi = 0, \pi/2, \pi, 3\pi/2$ as labeled. The outer envelope, which contains the range of coefficients that could be inferred from a sinusoid of constant amplitude but unknown phase, narrows as a larger fraction of an orbit is observed.}
    \label{fig:polycoeff}
\end{figure}

As observing duration becomes a significant fraction of an orbital period, parabolic fits to sinusoidal \ominusc\ data will depart from the results of the second-degree Taylor expansion described in Section~\ref{sec:analytic}. Instead of \ominusc\ measurements diverging from zero as time squared, a sinusoidal variation will turn back and generally flatten out, reducing the maximum coefficient we would fit for a parabola. In Figure~\ref{fig:polycoeff}, we show how the measured coefficient in front of the squared term of a parabolic fit evolves as the pulsating star is monitored for a larger fraction of an orbital period. Each gray curve represents how the inferred \pdot\ evolves as more data is collected for an intrinsically sinusoidal variation from a single-planet system for white dwarfs at different initial orbital phases at the start of observations. These are scaled relative to the upper limit on \pdot\ for that planet assuming edge-on ($i=\pi/2$) inclination (Eq.~\ref{eq:loggedO-C}). The black curves highlight initial phases $\phi_0 = 0, \pi/2, \pi, {\rm and\ } 3\pi/2$. When fitting parabolas to high-signal-to-noise sinusoidal data, the parabolic coefficient will evolve dramatically as more data are collected, even changing sign (e.g., Fig~\ref{fig:bigfig}), unlike expectations for measurements of an intrinsic \pdot, which will become more precise but should remain consistent as more data are collected. Significant departures to the best-fit coefficient as more data are collected is a signature that the \ominusc\ data are not parabolic in nature.

Overall the range of potential fit coefficients for any initial phase (bounded by Eq.~\ref{eq:loggedO-C}) decreases as the blue envelope containing all of these curves. The width of the uniform probability distribution for \pdot\ values that would be inferred from sinusoidal variations that are statistically indistinguishable from a parabola given the measurement noise is correspondingly reduced. 

The blue envelope in Figure~\ref{fig:polycoeff} also represents how the best-fit parabolic coefficient is reduced compared to the result for a Taylor limit expansion (Eq.~\ref{eq:physical}) for a constant orbital phase referenced at the midpoint of the observations (rather than the phase at the start of observations). This explains our results from a simulated \ominusc\ diagram for reflex motion in Section~\ref{sec:simulated} (Figure~\ref{fig:O-C}). In that example, the orbital phase at the midpoint of observations was set to $\phi=\frac{3\pi}{2}$, and 27\% of a full orbit was observed, resulting in an inferred \pdot\ around 5\% smaller than expected in the short-observing-duration limit.

\bibliography{references.bib}

\begin{thebibliography}{}
\expandafter\ifx\csname natexlab\endcsname\relax\def\natexlab#1{#1}\fi
\providecommand{\url}[1]{\href{#1}{#1}}
\providecommand{\dodoi}[1]{doi:~\href{http://doi.org/#1}{\nolinkurl{#1}}}
\providecommand{\doeprint}[1]{\href{http://ascl.net/#1}{\nolinkurl{http://ascl.net/#1}}}
\providecommand{\doarXiv}[1]{\href{https://arxiv.org/abs/#1}{\nolinkurl{https://arxiv.org/abs/#1}}}

\bibitem[{Alibert {et~al.}(2013)Alibert, Carron, Fortier, Pfyffer, Benz,
  Mordasini, \& Swoboda}]{Alibert_2013}
Alibert, Y., Carron, F., Fortier, A., {et~al.} 2013, Astronomy \& Astrophysics,
  558, A109, \dodoi{10.1051/0004-6361/201321690}

\bibitem[{Andryushin \& Popov(2021)}]{Andryushin_2021}
Andryushin, A.~S., \& Popov, S.~B. 2021, Astronomy Reports, 65, 246,
  \dodoi{10.1134/s1063772921040016}

\bibitem[{{Anguiano} {et~al.}(2017){Anguiano}, {Rebassa-Mansergas},
  {Garc{\'\i}a-Berro}, {Torres}, {Freeman}, \& {Zwitter}}]{2017MNRAS.469.2102A}
{Anguiano}, B., {Rebassa-Mansergas}, A., {Garc{\'\i}a-Berro}, E., {et~al.}
  2017, \mnras, 469, 2102, \dodoi{10.1093/mnras/stx796}

\bibitem[{{Bell}(2025)}]{2025arXiv250213258B}
{Bell}, K.~J. 2025, arXiv e-prints, arXiv:2502.13258,
  \dodoi{10.48550/arXiv.2502.13258}

\bibitem[{{Benvenuto} {et~al.}(2004){Benvenuto}, {Garc{\'\i}a-Berro}, \&
  {Isern}}]{gdot}
{Benvenuto}, O.~G., {Garc{\'\i}a-Berro}, E., \& {Isern}, J. 2004, \prd, 69,
  082002, \dodoi{10.1103/PhysRevD.69.082002}

\bibitem[{{Bischoff-Kim} {et~al.}(2008){Bischoff-Kim}, {Montgomery}, \&
  {Winget}}]{2008ApJ...675.1512B}
{Bischoff-Kim}, A., {Montgomery}, M.~H., \& {Winget}, D.~E. 2008, \apj, 675,
  1512, \dodoi{10.1086/526398}

\bibitem[{{Blackman} {et~al.}(2021){Blackman}, {Beaulieu}, {Bennett},
  {Danielski}, {Alard}, {Cole}, {Vandorou}, {Ranc}, {Terry}, {Bhattacharya},
  {Bond}, {Bachelet}, {Veras}, {Koshimoto}, {Batista}, \&
  {Marquette}}]{2021Natur.598..272B}
{Blackman}, J.~W., {Beaulieu}, J.~P., {Bennett}, D.~P., {et~al.} 2021, \nat,
  598, 272, \dodoi{10.1038/s41586-021-03869-6}

\bibitem[{Cassan {et~al.}(2012)Cassan, Kubas, Beaulieu, Dominik, Horne,
  Greenhill, Wambsganss, Menzies, Williams, J{\o}rgensen, Udalski, Bennett,
  Albrow, Batista, Brillant, Caldwell, Cole, Coutures, Cook, Dieters, Prester,
  Donatowicz, Fouqu{\'{e}}, Hill, Kains, Kane, Marquette, Martin, Pollard,
  Sahu, Vinter, Warren, Watson, Zub, Sumi, Szyma{\'{n}}ski, Kubiak, Poleski,
  Soszynski, Ulaczyk, Pietrzy{\'{n}}ski, \& Wyrzykowski}]{exoplanet}
Cassan, A., Kubas, D., Beaulieu, J.-P., {et~al.} 2012, Nature, 481, 167,
  \dodoi{10.1038/nature10684}

\bibitem[{{Childs} {et~al.}(2019){Childs}, {Quintana}, {Barclay}, \&
  {Steffen}}]{2019MNRAS.485..541C}
{Childs}, A.~C., {Quintana}, E., {Barclay}, T., \& {Steffen}, J.~H. 2019,
  \mnras, 485, 541, \dodoi{10.1093/mnras/stz385}

\bibitem[{C{\'{o}}rsico {et~al.}(2014)C{\'{o}}rsico, Althaus, Bertolami,
  Kepler, \& Garc{\'{\i}}a-Berro}]{neutrino}
C{\'{o}}rsico, A., Althaus, L., Bertolami, M.~M., Kepler, S., \&
  Garc{\'{\i}}a-Berro, E. 2014, Journal of Cosmology and Astroparticle Physics,
  2014, 054, \dodoi{10.1088/1475-7516/2014/08/054}

\bibitem[{{C{\'o}rsico} \& {Althaus}(2004)}]{2004A&A...428..159C}
{C{\'o}rsico}, A.~H., \& {Althaus}, L.~G. 2004, \aap, 428, 159,
  \dodoi{10.1051/0004-6361:20041372}

\bibitem[{{C{\'o}rsico} {et~al.}(2013){C{\'o}rsico}, {Althaus},
  {Garc{\'\i}a-Berro}, \& {Romero}}]{2013JCAP...06..032C}
{C{\'o}rsico}, A.~H., {Althaus}, L.~G., {Garc{\'\i}a-Berro}, E., \& {Romero},
  A.~D. 2013, \jcap, 2013, 032, \dodoi{10.1088/1475-7516/2013/06/032}

\bibitem[{{C{\'o}rsico} {et~al.}(2019){C{\'o}rsico}, {Althaus}, {Miller
  Bertolami}, \& {Kepler}}]{2019A&ARv..27....7C}
{C{\'o}rsico}, A.~H., {Althaus}, L.~G., {Miller Bertolami}, M.~M., \& {Kepler},
  S.~O. 2019, \aapr, 27, 7, \dodoi{10.1007/s00159-019-0118-4}

\bibitem[{{C{\'o}rsico} {et~al.}(2012{\natexlab{a}}){C{\'o}rsico}, {Althaus},
  {Miller Bertolami}, {Romero}, {Garc{\'\i}a-Berro}, {Isern}, \&
  {Kepler}}]{2012MNRAS.424.2792C}
{C{\'o}rsico}, A.~H., {Althaus}, L.~G., {Miller Bertolami}, M.~M., {et~al.}
  2012{\natexlab{a}}, \mnras, 424, 2792,
  \dodoi{10.1111/j.1365-2966.2012.21401.x}

\bibitem[{{C{\'o}rsico} {et~al.}(2012{\natexlab{b}}){C{\'o}rsico}, {Althaus},
  {Romero}, {Mukadam}, {Garc{\'\i}a-Berro}, {Isern}, {Kepler}, \&
  {Corti}}]{2012JCAP...12..010C}
{C{\'o}rsico}, A.~H., {Althaus}, L.~G., {Romero}, A.~D., {et~al.}
  2012{\natexlab{b}}, \jcap, 2012, 010, \dodoi{10.1088/1475-7516/2012/12/010}

\bibitem[{{C{\'o}rsico} {et~al.}(2016){C{\'o}rsico}, Romero, Althaus,
  García-Berro, Isern, Kepler, Bertolami, Sullivan, \& Chote}]{C_rsico_2016}
{C{\'o}rsico}, A.~H., Romero, A.~D., Althaus, L.~G., {et~al.} 2016, Journal of
  Cosmology and Astroparticle Physics, 2016, 036–036,
  \dodoi{10.1088/1475-7516/2016/07/036}

\bibitem[{Costa \& Kepler(2008)}]{Costa_2008}
Costa, J. E.~S., \& Kepler, S.~O. 2008, Astronomy \&; Astrophysics, 489,
  1225–1232, \dodoi{10.1051/0004-6361:20079118}

\bibitem[{{Dalessio} {et~al.}(2013{\natexlab{a}}){Dalessio}, {Provencal},
  {Barlow}, \& {Shipman}}]{2013ASPC..469...45D}
{Dalessio}, J., {Provencal}, J.~L., {Barlow}, B.~N., \& {Shipman}, H.~L.
  2013{\natexlab{a}}, in Astronomical Society of the Pacific Conference Series,
  Vol. 469, 18th European White Dwarf Workshop., ed. J.~{Krzesi{\'n}ski},
  G.~{Stachowski}, P.~{Moskalik}, \& K.~{Bajan}, 45

\bibitem[{{Dalessio} {et~al.}(2013{\natexlab{b}}){Dalessio}, {Sullivan},
  {Provencal}, {Shipman}, {Sullivan}, {Kilkenny}, {Fraga}, \&
  {Sefako}}]{2013ApJ...765....5D}
{Dalessio}, J., {Sullivan}, D.~J., {Provencal}, J.~L., {et~al.}
  2013{\natexlab{b}}, \apj, 765, 5, \dodoi{10.1088/0004-637X/765/1/5}

\bibitem[{{G{\"a}nsicke} {et~al.}(2019){G{\"a}nsicke}, {Schreiber}, {Toloza},
  {Gentile Fusillo}, {Koester}, \& {Manser}}]{2019Natur.576...61G}
{G{\"a}nsicke}, B.~T., {Schreiber}, M.~R., {Toloza}, O., {et~al.} 2019, \nat,
  576, 61, \dodoi{10.1038/s41586-019-1789-8}

\bibitem[{{Garc{\'\i}a-Berro} \& {Oswalt}(2016)}]{2016NewAR..72....1G}
{Garc{\'\i}a-Berro}, E., \& {Oswalt}, T.~D. 2016, \nar, 72, 1,
  \dodoi{10.1016/j.newar.2016.08.001}

\bibitem[{{Giammichele} {et~al.}(2022){Giammichele}, {Charpinet}, \&
  {Brassard}}]{2022FrASS...9.9045G}
{Giammichele}, N., {Charpinet}, S., \& {Brassard}, P. 2022, Frontiers in
  Astronomy and Space Sciences, 9, 879045, \dodoi{10.3389/fspas.2022.879045}

\bibitem[{{Guidry} {et~al.}(2021){Guidry}, {Vanderbosch}, {Hermes}, {Barlow},
  {Lopez}, {Boudreaux}, {Corcoran}, {Bell}, {Montgomery}, {Heintz},
  {Castanheira}, {Reding}, {Dunlap}, {Winget}, {Winget}, \&
  {Kuehne}}]{2021ApJ...912..125G}
{Guidry}, J.~A., {Vanderbosch}, Z.~P., {Hermes}, J.~J., {et~al.} 2021, \apj,
  912, 125, \dodoi{10.3847/1538-4357/abee68}

\bibitem[{{Hermes}(2018)}]{2018haex.bookE...6H}
{Hermes}, J.~J. 2018, in Handbook of Exoplanets, ed. H.~J. {Deeg} \& J.~A.
  {Belmonte}, 6, \dodoi{10.1007/978-3-319-55333-7_6}

\bibitem[{{Hermes} {et~al.}(2025){Hermes}, {Guidry}, {Vanderbosch},
  {Badenas-Agusti}, {Xu}, {Kao}, {Rodriguez}, \&
  {Hawkins}}]{2025ApJ...980...56H}
{Hermes}, J.~J., {Guidry}, J.~A., {Vanderbosch}, Z.~P., {et~al.} 2025, \apj,
  980, 56, \dodoi{10.3847/1538-4357/ada5fd}

\bibitem[{{Hermes} {et~al.}(2017){Hermes}, {G{\"a}nsicke}, {Kawaler}, {Greiss},
  {Tremblay}, {Gentile Fusillo}, {Raddi}, {Fanale}, {Bell}, {Dennihy}, {Fuchs},
  {Dunlap}, {Clemens}, {Montgomery}, {Winget}, {Chote}, {Marsh}, \&
  {Redfield}}]{Hermes2017}
{Hermes}, J.~J., {G{\"a}nsicke}, B.~T., {Kawaler}, S.~D., {et~al.} 2017, \apjs,
  232, 23, \dodoi{10.3847/1538-4365/aa8bb5}

\bibitem[{{Hsu} {et~al.}(2019){Hsu}, {Ford}, {Ragozzine}, \&
  {Ashby}}]{2019AJ....158..109H}
{Hsu}, D.~C., {Ford}, E.~B., {Ragozzine}, D., \& {Ashby}, K. 2019, \aj, 158,
  109, \dodoi{10.3847/1538-3881/ab31ab}

\bibitem[{{Isern} {et~al.}(1992){Isern}, {Hernanz}, \&
  {Garcia-Berro}}]{1992ApJ...392L..23I}
{Isern}, J., {Hernanz}, M., \& {Garcia-Berro}, E. 1992, \apjl, 392, L23,
  \dodoi{10.1086/186416}

\bibitem[{{Isern} {et~al.}(2022){Isern}, {Torres}, \&
  {Rebassa-Mansergas}}]{2022FrASS...9....6I}
{Isern}, J., {Torres}, S., \& {Rebassa-Mansergas}, A. 2022, Frontiers in
  Astronomy and Space Sciences, 9, 6, \dodoi{10.3389/fspas.2022.815517}

\bibitem[{{Kepler}(1993)}]{1993BaltA...2..515K}
{Kepler}, S.~O. 1993, Baltic Astronomy, 2, 515,
  \dodoi{10.1515/astro-1993-3-425}

\bibitem[{{Kepler} {et~al.}(1991){Kepler}, {Winget}, {Nather}, {Bradley},
  {Grauer}, {Fontaine}, {Bergeron}, {Vauclair}, {Claver}, {Marar}, {Seetha},
  {Ashoka}, {Mazeh}, {Leibowitz}, {Dolez}, {Chevreton}, {Barstow}, {Clemens},
  {Kleinman}, {Sansom}, {Tweedy}, {Kanaan}, {Hine}, {Provencal}, {Wesemael},
  {Wood}, {Brassard}, {Solheim}, \& {Emanuelsen}}]{1991ApJ...378L..45K}
{Kepler}, S.~O., {Winget}, D.~E., {Nather}, R.~E., {et~al.} 1991, \apjl, 378,
  L45, \dodoi{10.1086/186138}

\bibitem[{Kepler {et~al.}(2020)Kepler, Winget, Vanderbosch, Castanheira,
  Hermes, Bell, Mullally, Romero, Montgomery, DeGennaro, Winget, Chandler,
  Jeffery, Fritzen, Williams, Chote, \& Zola}]{stableoptical}
Kepler, S.~O., Winget, D.~E., Vanderbosch, Z.~P., {et~al.} 2020, The
  Astrophysical Journal, 906, 7, \dodoi{10.3847/1538-4357/abc626}

\bibitem[{{Kipping}(2024)}]{2024MNRAS.527.3532K}
{Kipping}, D. 2024, \mnras, 527, 3532, \dodoi{10.1093/mnras/stad3431}

\bibitem[{{Luhman} {et~al.}(2011){Luhman}, {Burgasser}, \&
  {Bochanski}}]{2011ApJ...730L...9L}
{Luhman}, K.~L., {Burgasser}, A.~J., \& {Bochanski}, J.~J. 2011, \apjl, 730,
  L9, \dodoi{10.1088/2041-8205/730/1/L9}

\bibitem[{{Manser} {et~al.}(2019){Manser}, {G{\"a}nsicke}, {Eggl}, {Hollands},
  {Izquierdo}, {Koester}, {Landstreet}, {Lyra}, {Marsh}, {Meru}, {Mustill},
  {Rodr{\'\i}guez-Gil}, {Toloza}, {Veras}, {Wilson}, {Burleigh}, {Davies},
  {Farihi}, {Gentile Fusillo}, {de Martino}, {Parsons}, {Quirrenbach}, {Raddi},
  {Reffert}, {Del Santo}, {Schreiber}, {Silvotti}, {Toonen}, {Villaver},
  {Wyatt}, {Xu}, \& {Portegies Zwart}}]{2019Sci...364...66M}
{Manser}, C.~J., {G{\"a}nsicke}, B.~T., {Eggl}, S., {et~al.} 2019, Science,
  364, 66, \dodoi{10.1126/science.aat5330}

\bibitem[{{Montgomery} {et~al.}(2020){Montgomery}, {Hermes}, {Winget},
  {Dunlap}, \& {Bell}}]{Montgomery2020}
{Montgomery}, M.~H., {Hermes}, J.~J., {Winget}, D.~E., {Dunlap}, B.~H., \&
  {Bell}, K.~J. 2020, \apj, 890, 11, \dodoi{10.3847/1538-4357/ab6a0e}

\bibitem[{{Mukadam} {et~al.}(2013){Mukadam}, {Bischoff-Kim}, {Fraser},
  {C{\'o}rsico}, {Montgomery}, {Kepler}, {Romero}, {Winget}, {Hermes},
  {Riecken}, {Kronberg}, {Winget}, {Falcon}, {Chandler}, {Kuehne}, {Sullivan},
  {Reaves}, {von Hippel}, {Mullally}, {Shipman}, {Thompson}, {Silvestri}, \&
  {Hynes}}]{R548}
{Mukadam}, A.~S., {Bischoff-Kim}, A., {Fraser}, O., {et~al.} 2013, \apj, 771,
  17, \dodoi{10.1088/0004-637X/771/1/17}

\bibitem[{Mullally {et~al.}(2008)Mullally, Winget, Degennaro, Jeffery,
  Thompson, Chandler, \& Kepler}]{mullally}
Mullally, F., Winget, D.~E., Degennaro, S., {et~al.} 2008, The Astrophysical
  Journal, 676, 573, \dodoi{10.1086/528672}

\bibitem[{{Mullally} {et~al.}(2024){Mullally}, {Debes}, {Cracraft}, {Mullally},
  {Poulsen}, {Albert}, {Thibault}, {Reach}, {Hermes}, {Barclay}, {Kilic}, \&
  {Quintana}}]{2024ApJ...962L..32M}
{Mullally}, S.~E., {Debes}, J., {Cracraft}, M., {et~al.} 2024, \apjl, 962, L32,
  \dodoi{10.3847/2041-8213/ad2348}

\bibitem[{{Murphy} {et~al.}(2014){Murphy}, {Bedding}, {Shibahashi}, {Kurtz}, \&
  {Kjeldsen}}]{2014MNRAS.441.2515M}
{Murphy}, S.~J., {Bedding}, T.~R., {Shibahashi}, H., {Kurtz}, D.~W., \&
  {Kjeldsen}, H. 2014, \mnras, 441, 2515, \dodoi{10.1093/mnras/stu765}

\bibitem[{{Musielak} {et~al.}(2005){Musielak}, {Cuntz}, {Marshall}, \&
  {Stuit}}]{2005A&A...434..355M}
{Musielak}, Z.~E., {Cuntz}, M., {Marshall}, E.~A., \& {Stuit}, T.~D. 2005,
  \aap, 434, 355, \dodoi{10.1051/0004-6361:20040238}

\bibitem[{{Niu} {et~al.}(2018){Niu}, {Li}, {Zong}, {Xue}, \&
  {Wang}}]{2018PhRvD..98j3023N}
{Niu}, J.-S., {Li}, T., {Zong}, W., {Xue}, H.-F., \& {Wang}, Y. 2018, \prd, 98,
  103023, \dodoi{10.1103/PhysRevD.98.103023}

\bibitem[{{Oliveira da Rosa} {et~al.}(2022){Oliveira da Rosa}, {Kepler},
  {C{\'o}rsico}, {Costa}, {Hermes}, {Kawaler}, {Bell}, {Montgomery},
  {Provencal}, {Winget}, {Handler}, {Dunlap}, {Clemens}, \&
  {Uzundag}}]{2022ApJ...936..187O}
{Oliveira da Rosa}, G., {Kepler}, S.~O., {C{\'o}rsico}, A.~H., {et~al.} 2022,
  \apj, 936, 187, \dodoi{10.3847/1538-4357/ac8871}

\bibitem[{{Paxton} {et~al.}(2011){Paxton}, {Bildsten}, {Dotter}, {Herwig},
  {Lesaffre}, \& {Timmes}}]{MESA}
{Paxton}, B., {Bildsten}, L., {Dotter}, A., {et~al.} 2011, \apjs, 192, 3,
  \dodoi{10.1088/0067-0049/192/1/3}

\bibitem[{{Redaelli} {et~al.}(2011){Redaelli}, {Kepler}, {Costa}, {Winget},
  {Handler}, {Castanheira}, {Kanaan}, {Fraga}, {Henrique}, {Giovannini},
  {Provencal}, {Shipman}, {Dalessio}, {Thompson}, {Mullally}, {Brewer},
  {Childers}, {Oksala}, {Rosen}, {Wood}, {Reed}, {Walter}, {Strickland},
  {Chandler}, {Watson}, {Nather}, {Montgomery}, {Bischoff-Kim}, {Hansen},
  {Nitta}, {Kleinman}, {Claver}, {Brown}, {Sullivan}, {Kim}, {Chen}, {Yang},
  {Shih}, {Zhang}, {Jiang}, {Fu}, {Seetha}, {Ashoka}, {Marar}, {Baliyan},
  {Vats}, {Chernyshev}, {Ibbetson}, {Leibowitz}, {Hemar}, {Sergeev}, {Andreev},
  {Janulis}, {Mei{\v{s}}tas}, {Moskalik}, {Pajdosz}, {Baran}, {Winiarski},
  {Zola}, {Ogloza}, {Siwak}, {Bogn{\'a}r}, {Solheim}, {Sefako}, {Buckley},
  {O'Donoghue}, {Nagel}, {Silvotti}, {Bruni}, {Fremy}, {Vauclair}, {Chevreton},
  {Dolez}, {Pfeiffer}, {Barstow}, {Creevey}, {Kawaler}, \& {Clemens}}]{rad}
{Redaelli}, M., {Kepler}, S.~O., {Costa}, J.~E.~S., {et~al.} 2011, \mnras, 415,
  1220, \dodoi{10.1111/j.1365-2966.2011.18743.x}

\bibitem[{Robinson(2016)}]{textbook}
Robinson, E.~L. 2016, Data Analysis for Scientist and Engineers (Princeton
  University Press)

\bibitem[{{Sigurdsson} {et~al.}(2003){Sigurdsson}, {Richer}, {Hansen},
  {Stairs}, \& {Thorsett}}]{2003Sci...301..193S}
{Sigurdsson}, S., {Richer}, H.~B., {Hansen}, B.~M., {Stairs}, I.~H., \&
  {Thorsett}, S.~E. 2003, Science, 301, 193, \dodoi{10.1126/science.1086326}

\bibitem[{{Tremblay} {et~al.}(2019){Tremblay}, {Fontaine}, {Gentile Fusillo},
  {Dunlap}, {G{\"a}nsicke}, {Hollands}, {Hermes}, {Marsh}, {Cukanovaite}, \&
  {Cunningham}}]{2019Natur.565..202T}
{Tremblay}, P.-E., {Fontaine}, G., {Gentile Fusillo}, N.~P., {et~al.} 2019,
  \nat, 565, 202, \dodoi{10.1038/s41586-018-0791-x}

\bibitem[{{van Horn}(1968)}]{1968ApJ...151..227V}
{van Horn}, H.~M. 1968, \apj, 151, 227, \dodoi{10.1086/149432}

\bibitem[{{Vanderbosch} {et~al.}(2021){Vanderbosch}, {Rappaport}, {Guidry},
  {Gary}, {Blouin}, {Kaye}, {Weinberger}, {Melis}, {Klein}, {Zuckerman},
  {Vanderburg}, {Hermes}, {Hegedus}, {Burleigh}, {Sefako}, {Worters}, \&
  {Heintz}}]{2021ApJ...917...41V}
{Vanderbosch}, Z.~P., {Rappaport}, S., {Guidry}, J.~A., {et~al.} 2021, \apj,
  917, 41, \dodoi{10.3847/1538-4357/ac0822}

\bibitem[{{Vanderburg} {et~al.}(2015){Vanderburg}, {Johnson}, {Rappaport},
  {Bieryla}, {Irwin}, {Lewis}, {Kipping}, {Brown}, {Dufour}, {Ciardi}, {Angus},
  {Schaefer}, {Latham}, {Charbonneau}, {Beichman}, {Eastman}, {McCrady},
  {Wittenmyer}, \& {Wright}}]{2015Natur.526..546V}
{Vanderburg}, A., {Johnson}, J.~A., {Rappaport}, S., {et~al.} 2015, \nat, 526,
  546, \dodoi{10.1038/nature15527}

\bibitem[{{Vanderburg} {et~al.}(2020){Vanderburg}, {Rappaport}, {Xu},
  {Crossfield}, {Becker}, {Gary}, {Murgas}, {Blouin}, {Kaye}, {Palle}, {Melis},
  {Morris}, {Kreidberg}, {Gorjian}, {Morley}, {Mann}, {Parviainen}, {Pearce},
  {Newton}, {Carrillo}, {Zuckerman}, {Nelson}, {Zeimann}, {Brown},
  {Tronsgaard}, {Klein}, {Ricker}, {Vanderspek}, {Latham}, {Seager}, {Winn},
  {Jenkins}, {Adams}, {Benneke}, {Berardo}, {Buchhave}, {Caldwell},
  {Christiansen}, {Collins}, {Col{\'o}n}, {Daylan}, {Doty}, {Doyle},
  {Dragomir}, {Dressing}, {Dufour}, {Fukui}, {Glidden}, {Guerrero}, {Guo},
  {Heng}, {Henriksen}, {Huang}, {Kaltenegger}, {Kane}, {Lewis}, {Lissauer},
  {Morales}, {Narita}, {Pepper}, {Rose}, {Smith}, {Stassun}, \&
  {Yu}}]{2020Natur.585..363V}
{Vanderburg}, A., {Rappaport}, S.~A., {Xu}, S., {et~al.} 2020, \nat, 585, 363,
  \dodoi{10.1038/s41586-020-2713-y}

\bibitem[{{Veras} \& {G{\"a}nsicke}(2015)}]{2015MNRAS.447.1049V}
{Veras}, D., \& {G{\"a}nsicke}, B.~T. 2015, \mnras, 447, 1049,
  \dodoi{10.1093/mnras/stu2475}

\bibitem[{{Villaver} \& {Livio}(2009)}]{2009ApJ...705L..81V}
{Villaver}, E., \& {Livio}, M. 2009, \apjl, 705, L81,
  \dodoi{10.1088/0004-637X/705/1/L81}

\bibitem[{{Williams} {et~al.}(2009){Williams}, {Bolte}, \&
  {Koester}}]{2009ApJ...693..355W}
{Williams}, K.~A., {Bolte}, M., \& {Koester}, D. 2009, \apj, 693, 355,
  \dodoi{10.1088/0004-637X/693/1/355}

\bibitem[{{Winget} \& {Kepler}(2008)}]{2008ARA&A..46..157W}
{Winget}, D.~E., \& {Kepler}, S.~O. 2008, \araa, 46, 157,
  \dodoi{10.1146/annurev.astro.46.060407.145250}

\bibitem[{{Winget} {et~al.}(1985){Winget}, {Kepler}, {Robinson}, {Nather}, \&
  {Odonoghue}}]{1985ApJ...292..606W}
{Winget}, D.~E., {Kepler}, S.~O., {Robinson}, E.~L., {Nather}, R.~E., \&
  {Odonoghue}, D. 1985, \apj, 292, 606, \dodoi{10.1086/163193}

\bibitem[{{Winget} {et~al.}(2015){Winget}, {Hermes}, {Mullally}, {Bell},
  {Montgomery}, {Williams}, {Harrold}, {Kepler}, {Castanheira}, {Chandler},
  {Winget}, {Mukadam}, \& {Nather}}]{2015ASPC..493..285W}
{Winget}, D.~E., {Hermes}, J.~J., {Mullally}, F., {et~al.} 2015, in
  Astronomical Society of the Pacific Conference Series, Vol. 493, 19th
  European Workshop on White Dwarfs, ed. P.~{Dufour}, P.~{Bergeron}, \&
  G.~{Fontaine}, 285

\bibitem[{{Yao}(2023)}]{msthesis}
{Yao}, L.~X. 2023, Master's thesis, CUNY Queens College, Queens, NY

\end{thebibliography}

\end{document}